# Elastic Rod Origami (RodOri)
# for Programming Static and Dynamic Mechanical Properties

**Authors:**

Sophie Leanza,[1]† Jeseung Lee,[1]† Ruike Renee Zhao[1]*

**Affiliations:**

[1]Stanford University; Stanford, CA 94305, USA

*Corresponding author. Email: rrzhao@stanford.edu

†These authors contributed equally to this work.

**Abstract:**

Reconfigurable mechanical systems enable precise programmable control over structural properties, opening new opportunities in architected materials, adaptive devices, and multifunctional structures. Here, we introduce elastic rod origami (RodOri), a platform that exploits remarkably simple elements—pre-stressed, naturally curved rods—into a system with an extraordinary degree of multistability and configurational richness. For example, a single 6-rod RodOri unit can easily access 11 distinct configurations, far exceeding the reconfigurability of conventional origami or general mechanical reconfigurable systems. Individual rods, constrained under clamped boundary conditions, undergo snapping transitions between discrete morphologies whose strain energy and stiffness are precisely prescribed by their natural curvature. Assembling these rods into modular multi-rod architectures yields metamaterials with numerous stable configurations that can be selectively and reversibly programmed. This configurational diversity enables tunable static stiffness, nonlinear force response, and thus enables tunable dynamic behaviors such as vibration filtering, wave-propagation switching, and mode conversion within a single, easily-manufactured platform. By leveraging curvature-induced mechanical instability, RodOri unlocks highly programmable static and dynamic mechanical behavior, offering new design strategies in reconfigurable structures, soft robotics, medical devices, and adaptive materials.

**Main Text:**

**1. Introduction**

Reconfigurable mechanical systems [1-6] are rapidly transforming the design of architected materials and structures by enabling tunable or programmable properties and functions such as stiffness [1, 2, 5, 7], strength [1, 6], mechanical logic/memory [8-10], and dynamic response such as wave manipulation [11-14]. Among various reconfiguration strategies, mechanical instability offers a particularly powerful mechanism [15] that allows systems to access multiple stable states, store and release elastic energy, or undergo rapid shape transitions in response to external stimuli. Snap buckling, in particular, is advantageous for achieving fast, reversible transformations and has been widely harnessed in soft robotics [16], metamaterials [17], and diverse multistable structures [18-20]. By leveraging instability in beams, rods, plates, and shells, engineers have developed mechanical systems for energy absorption [21-23] and storage [24], vibration control [25, 26], and logic functions [9, 27, 28].

Here, we introduce elastic rod origami (RodOri), a fundamentally new class of reconfigurable mechanical systems, by using elastic instabilities in pre-stressed rod elements to achieve programmable multistability and multifunctionality. Unlike conventional approaches that achieve reconfigurability through bistable elements [1, 29-32] and/or hinge-based linkages [13, 14, 33-35], RodOri exploits geometrically simple yet versatile curved rods to encode elastic instabilities that drive a set of discrete, reconfigurable states within a multistable architecture. Crucially, the diverse elastic energy landscapes associated with these multiple stable states enable programmable mechanical properties, both static and dynamic, empowering the structure to perform diverse, adaptive functions and seamlessly switch between them on demand.

At the core of this platform is the pre-stressed rod constrained by clamped boundary conditions, referred to as the Hutchinson rod [36], which acts as the fundamental element of the RodOri. When straightened and clamped at both ends and subjected to axial compression, the Hutchinson rod undergoes a characteristic snap-buckling transition (**Fig. 1a**). This snapping response is governed by the rod's intrinsic natural curvature ($\kappa_n$) in its stress-free state, which defines its sole stable state (i.e., monostable) in isolation. Notably, when two such rods are straightened by applying opposing external bending moments and rigidly joined at their ends, they form a self-equilibrated "bi-rod" structure (**Fig. 1b**), in which the internal bending moments counterbalance one another, yielding a stable bi-rod. Upon axial compression, the bi-rod snaps into a looped configuration, resulting in a second mechanically stable state. Furthermore, a third stable state emerges when bending is applied at the ends, illustrating the intrinsic multistability of the bi-rod unit.

Building on the bi-rod concept, larger assemblies of Hutchinson rods can be architected into RodOri structures with multiple mechanically stable configurations. These structures can reversibly transition between states through elastic deformation-induced snapping, enabling mechanical programmability and reconfigurability. **Fig. 1c** shows an example of such a structure composed of six rods. While stable in its straightened form (deployed), the assembly can be axially compressed into a second stable state (orange, 0° twist) or rotated in discrete 60° increments to achieve additional stable states, totaling six distinct configurations. More configurations are possible by changing the rod geometry. We refer to this multi-rod construct as a single RodOri unit. Both the Hutchinson rods and rigid bases for RodOri are 3D-printed, making them simple to produce and easily adaptable to different designs. By stacking or tessellating such units (**Fig. 1d**), we realize a modular platform for selective reconfiguration, allowing mechanical properties and thus functionalities to be tailored across different global states. Owing to the intrinsic reconfigurability ($n$ configurations) of each unit, independently programming the state of each constituent within an assembly made of $m$ units unlocks a combinatorially rich and immense design space of $n^m$ possible global configurations, enabling fine-grained programmability of global static and dynamic mechanical properties and, in turn, tunable functionalities across the entire structure.

In this work, we investigate the snapping behavior of individual pre-stressed rods with varied natural curvatures and systematically characterize the multistable states of RodOri achieved through twisting, along with the strain energy landscape of the structure. Remarkably, an individual RodOri unit can have as many as eleven discrete configurations. Stacked assemblies of these units (**Fig. 1d**, left) exhibit tunable nonlinear stiffnesses through controlled transitions between configurations. Extending this concept, we construct active metamaterials in the form of tessellations of reconfigurable RodOri units (**Fig. 1d**, right), demonstrating programmable vibration control, including vibration isolation, amplification, and mode conversion. Considering the multiple configurations of RodOri and the number of units in an assembly, selective reconfiguration enables a vast range of architectural combinations with readily adaptable functionalities. Together, these results establish RodOri as a platform for easy-to-manufacture architected structures with mechanically programmable stiffness and dynamic response, offering a broadly applicable framework for harnessing mechanical instabilities in multifunctional and reconfigurable systems.

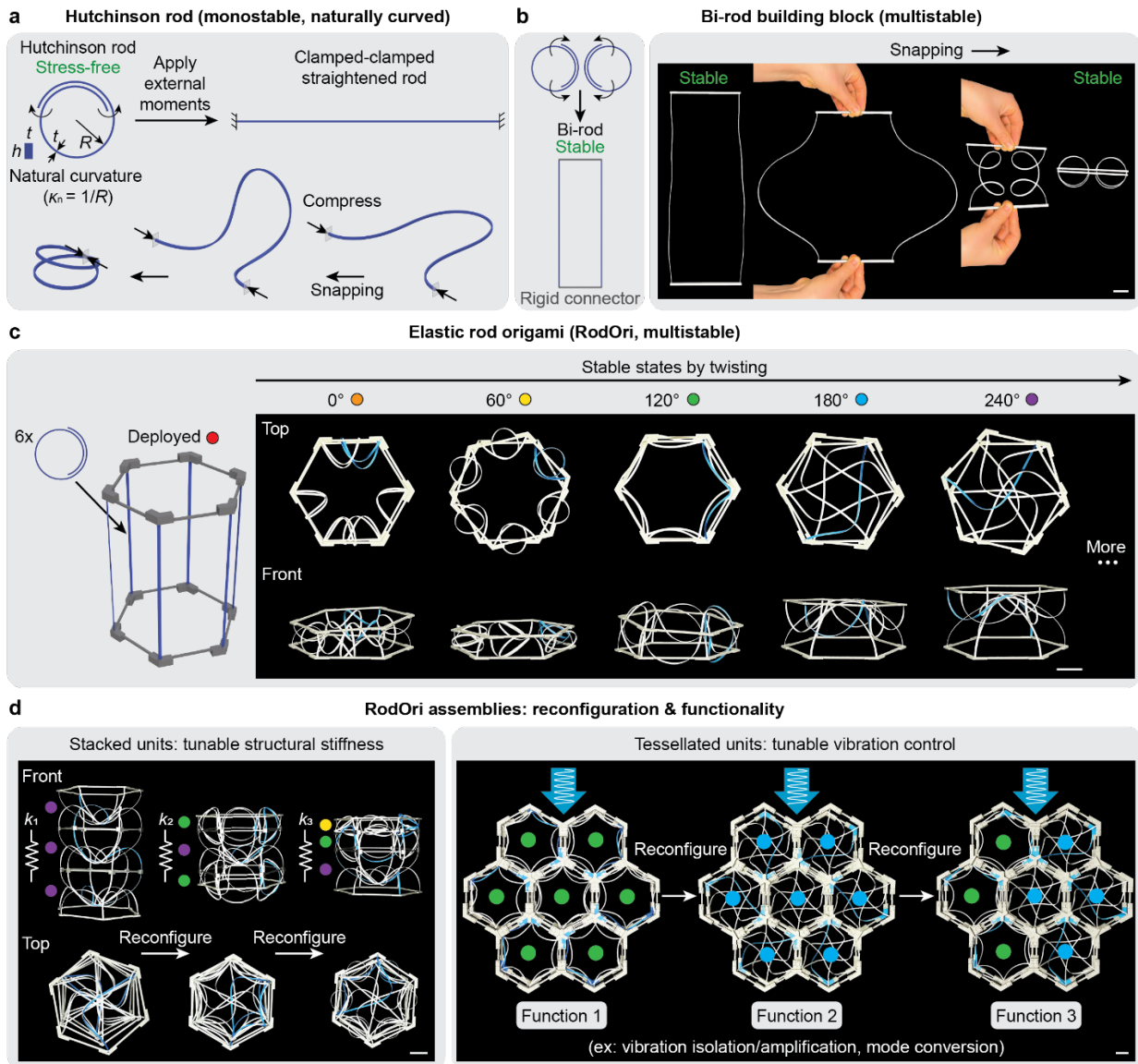

**Fig. 1 | Elastic rod origami (RodOri). a**, Naturally curved rod, termed a Hutchinson rod, which snap-buckles in 3D when compressed axially after straightening. **b**, Bi-rod structure, stable at the straightened state due to the balance of internal moments. When compressed axially, the bi-rod snaps to a looped second stable state. **c**, RodOri unit assembled from six straightened rods,

achieving several stable folded configurations which can be switched between by twisting. **d**, Assemblies of stacked and tessellated RodOri and their functionality. Tunable nonlinear stiffness is enabled by controlled transitions between global configurations. Individual units of a RodOri tessellation can be independently reconfigured for fine-grained programmability of global dynamic mechanical properties and tunable vibration control, including vibration isolation, amplification, or mode conversion. Scale bars: 2.5 cm.

## 2. Mechanical Behavior of Hutchinson Rods and RodOri

To understand the mechanical response and functional potential of the RodOri, we begin by investigating the axial compression behavior of a single Hutchinson rod [36]. Unlike traditional buckling elements, the Hutchinson rod offers a rich, designable mechanical response, arising from its intrinsic natural curvature and slender geometry that enables snap-through-driven shape transformations. Its behavior is governed by a set of parameters, including its length, $L$, stress-free radius of curvature $R$ (and thus its natural curvature $\kappa_n = 1/R$), and cross-sectional dimensions, height $h$ and thickness $t$. For all demonstrations in this work, we fix the aspect ratio to $h/t = 4$. Upon clamping both ends and compressing the rod, it buckles, with the axial end shortening characterized by normalized displacement $\Delta/L$ (**Fig. 2a**). Uniquely, the rod snaps when its dimensionless natural curvature $L\kappa_n/2\pi$ exceeds a threshold value, revealing a behavior fundamentally distinct from classic Euler buckling. For rods with $L = 200$ mm and varied $L\kappa_n/2\pi$ of 0.50, 1.27, 1.75, and 2.00, all initially buckle in-plane but the rods with higher natural curvature ($L\kappa_n/2\pi$ of 1.27, 1.75, and 2.00) then undergo a dramatic snap-through to a twisted, out-of-plane state as compression increases. Larger $\kappa_n$ leads to an earlier snapping point, serving as a way to design the rod's instability.

The snapping behavior, programmable via $L\kappa_n/2\pi$, introduces a novel design strategy in elastic structures, enabling tunable functional post-buckling responses through geometric prescription. The force-displacement relationship, both in the loading and unloading curves after snapping, implies that the rod's snapped configuration behaves as an effective elastic spring whose post-snapping stiffness $k$ is continuously tunable via $\kappa_n$, as shown in **Fig. 2a**. Upon unloading (dashed line in the plot of **Fig. 2a**), the rod snaps back to its straight configuration, following a distinct path and occurring at a small end shortening ($\Delta/L<0.05$). This hysteretic response further demonstrates the buckled rod's capacity to effectively function as a rod-based spring across a broad displacement range ($0.2 \lesssim \Delta/L \leq 1$). This large operational window is particularly advantageous for vibration-related applications, where consistent and tunable mechanical response under large-amplitude oscillations is critical. This tunable, reversible snapping behavior forms the foundation for the programmable mechanical properties of RodOri and is later exploited for dynamic stiffness modulation and vibration control.

To enable standalone reconfigurable functionality, multiple rods can be assembled into self-equilibrated structures by rigidly connecting their ends at polygonal bases, such that their internal bending moments counterbalance, resulting in a multi-rod RodOri unit (as introduced earlier). Here, we investigate a 6-rod RodOri unit subjected to axial compression. For assemblies composed of rods with low natural curvature (e.g., $L\kappa_n/2\pi = 0.50$) (**Fig. 2b**, left), the rods exhibit extended in-plane buckling followed by eventual out-of-plane deformation at large end shortening. The strain energy rises monotonically during loading (red curve in **Fig. 2c**), and the structure fully rebounds to its initial deployed configuration upon unloading, demonstrating monostable behavior. In contrast, for higher natural curvatures (e.g., $L\kappa_n/2\pi = 1.27$) (**Fig. 2b**, right), the rods snap under compression, collapsing into a stable folded configuration that persists after removing the compression, indicating the potential for multistability. The FEA (finite element analysis) (bottom row of **Fig. 2b**) closely replicates the experimental results, as well as captures the associated strain energy evolution (**Fig. 2c**).

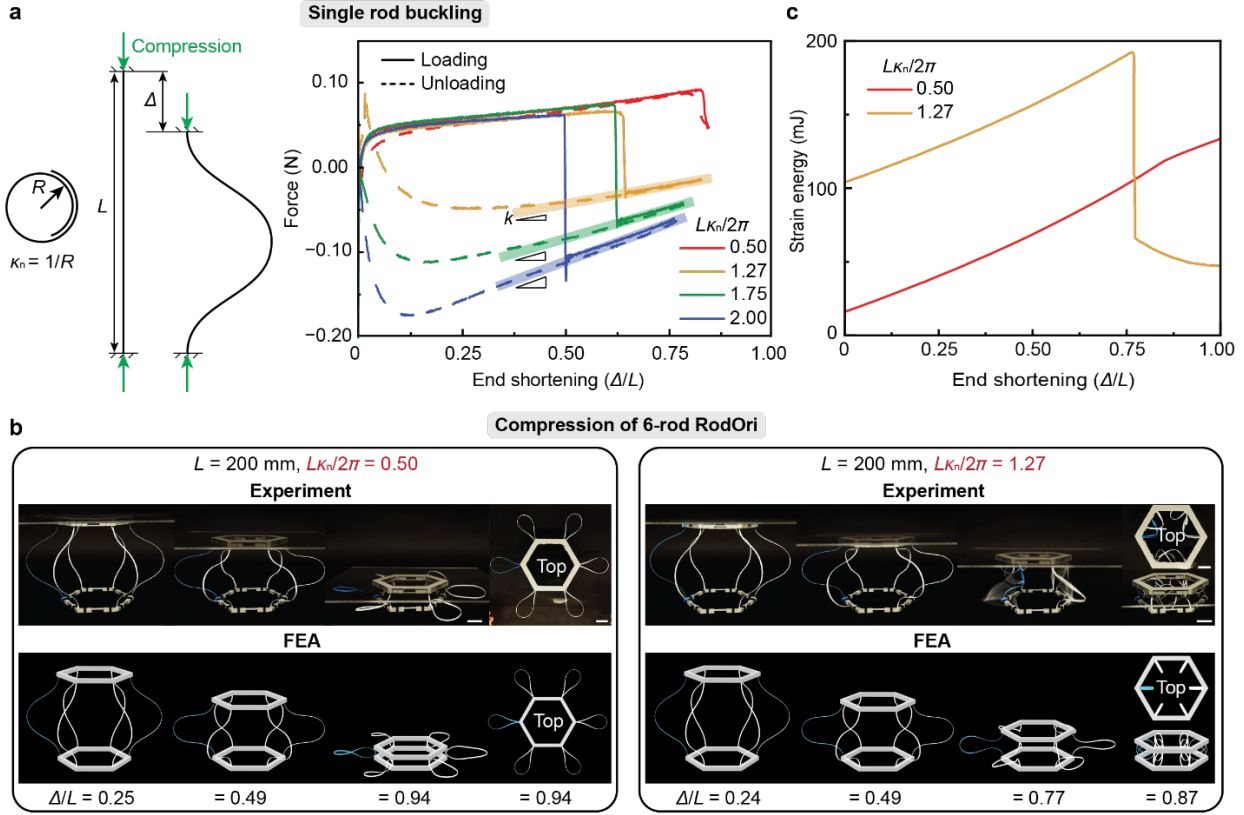

**Fig. 2 | Mechanical behavior of Hutchinson rod and RodOri. a**, Compression of Hutchinson rod, with its geometric parameters defined. Loading and unloading curves of the compressive force vs end shortening ($\Delta/L$) for rods with fixed $h/t = 4$ and $L = 200$ mm, and varied $L\kappa_n/2\pi$ of 0.50, 1.27, 1.75, and 2.00. **b**, Experimental and FEA results of a 6-rod RodOri unit with $h/t = 4$, $L = 200$ mm, and varied $L\kappa_n/2\pi$ of 0.50 and 1.27 under compression by a clear plate. **c**, Corresponding FEA results of the strain energy during compression of the 6-rod RodOri unit. Scale bars: 2.5 cm.

## 3. Multiple Configurations and Multistability of RodOri

Beyond compression, a far richer set of stable configurations can be accessed through twisting, enabling programmable control over both the static and dynamic mechanical properties of RodOri. Here, we explore the multiple configurations, multistability, and the associated elastic energy landscapes of a 6-rod RodOri unit. As shown in **Fig. 3a**, twisting the top base by an angle $\varphi$ while allowing vertical displacement leads to discrete, stable configurations. For rods with $h/t = 4$, $L = 200$ mm, and $L\kappa_n/2\pi = 1.27$, the structure can be twisted up to 240° (upper limit of twist due to contact between adjacent rods), resulting in six distinct stable states: deployed (all rods are straightened, no twist), folded (0°), and intermediate stable configurations at 60°, 120°, 180°, and 240° (**Fig. 3b**). Increasing the rod length to $L = 300$ mm ($L\kappa_n/2\pi = 1.91$) allows for further twisting to $\varphi = 540°$ (**Fig. 3c**), where eleven discrete configurations emerge at 60° intervals (**Fig. 3d**).

Among these, eight are stable standalone states (deployed, 0°, 60°, 180°, 240°, 300°, 360°, and 540°). Since each RodOri unit possesses $n$ reliably addressable configurations, it can in principle be viewed as an $n$-ary (in contrast to binary) logic element whose broad state space could enable adaptive signal routing and mechanical memory in reconfigurable metamaterials.

Each configuration is associated with a distinct level of stored elastic energy, which varies significantly with rod natural curvature (see **Fig. 3e** for configuration and strain energy from FEA). As shown in **Fig. 3e**, the energy landscape (represented by discrete points) can be precisely programmed by tailoring $L\kappa_n/2\pi$ and reconfiguring the unit into different states. This energy landscape, combined with the morphology-induced interactions among the rods, governs the unit's mechanical response to external loads, enabling fine control over both stiffness and dynamic behavior. Notably, units with different natural curvatures adopt distinct folded morphologies at the same twist angle (e.g., configuration at $\varphi = 120°$ in **Fig. 3b** and **Fig. 3d**), leading to markedly different strain energy landscapes. This natural curvature-induced control over both structural morphology and energy provides a powerful means for property tuning.

When units with the same rod natural curvatures are stacked or tessellated, for example, those with $L\kappa_n/2\pi = 1.91$, all eleven discrete configurations can serve as programmable building blocks where each unit can be independently programmed. This yields a combinatorial design space with up to $11^m$ possible global configurations and energy distributions for an array of $m$ units. Even a small tessellation of seven units (e.g., **Fig. 1d**) enables nearly 20 million unique combinations of structural states and energy distributions. To expand the design space further, units with different natural curvatures can be tessellated with one another to achieve heterogeneous assemblies capable of fine-tuning the structure's spatially varying mechanical properties and responses.

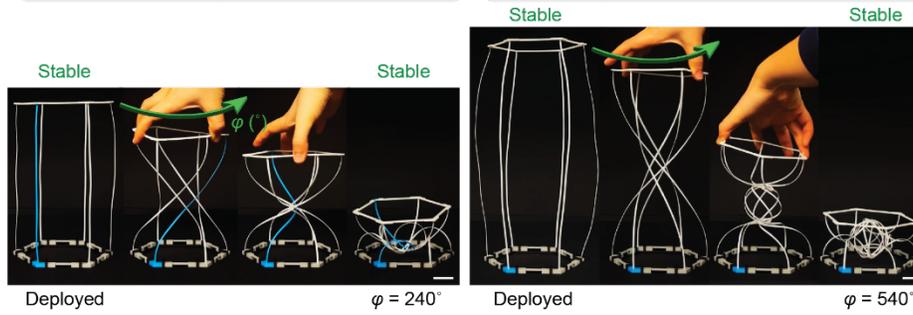

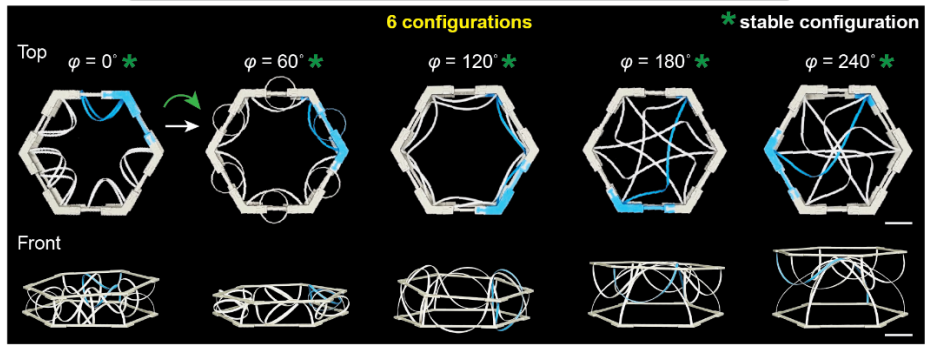

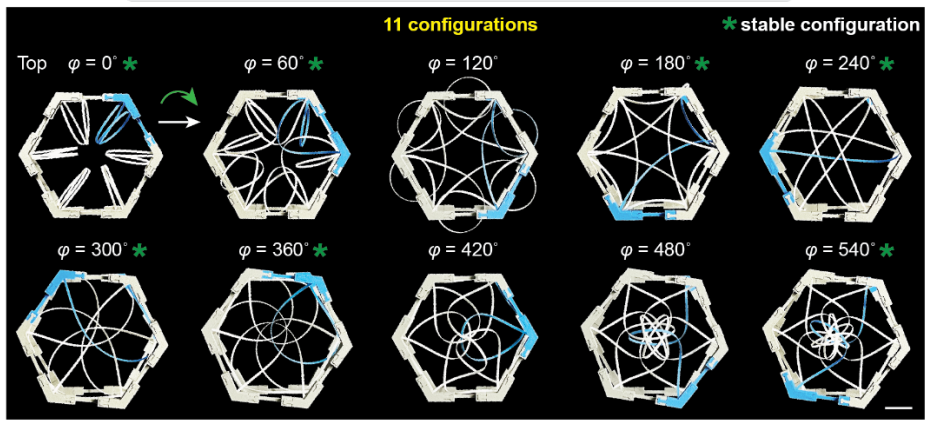

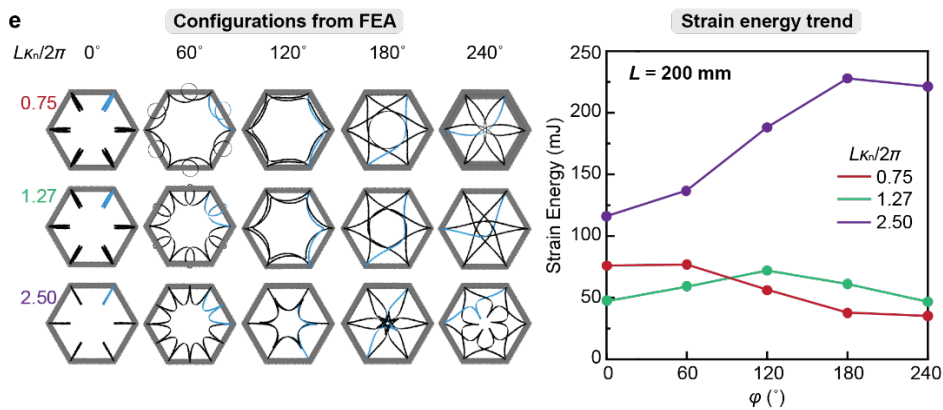

**Fig. 3 | Multiple configurations and multistability of RodOri. a**, Reconfiguration of 6-rod RodOri ($L = 200$ mm, $L\kappa_n/2\pi = 1.27$, $h/t = 4$) to 240° stable state. **b**, Top and front views of the

reconfiguration of 6-rod RodOri ($L = 200$ mm, $L\kappa_n/2\pi = 1.27$, $h/t = 4$) between five different stable states. Each state can be switched between by twisting the top base by 60°, ranging between 0° and 240° states. **c**, Reconfiguration of 6-rod RodOri ($L = 300$ mm, $L\kappa_n/2\pi = 1.91$, $h/t = 4$) to 540° stable state. **d**, Top view of the reconfiguration of 6-rod RodOri ($L = 300$ mm, $L\kappa_n/2\pi = 1.91$, $h/t = 4$) between ten different configurations, ranging from 0° and 540° in 60° increments. **e**, FEA results of structural configurations and the stored strain energy for the 6-rod RodOri (fixed $L = 200$ mm, $h/t = 4$) with different $L\kappa_n/2\pi$ of 0.75, 1.27, and 2.50. To the right, the discrete strain energy landscape at different angles of twist is shown for the different $L\kappa_n/2\pi$. Scale bars: 2.5 cm.

## 4. Static Stiffness Tuning of Stacked RodOri

Building on the programmable multistability and energy landscapes of RodOri, we now demonstrate how geometric reconfiguration enables broad and tunable control of axial stiffness. As the rods transition between discrete 3D configurations, their stored strain energy and mechanical response change, making shape morphing a direct mechanism for tailoring structural stiffness. Here, we measure the axial stiffness of the 6-rod RodOri unit ($L = 200$ mm, $L\kappa_n/2\pi = 1.27$, $h/t = 4$) across its five stable folded configurations (0°, 60°, 120°, 180°, 240°) through tensile testing, with axial end extension characterized by normalized displacement ($\Delta/L$). The stiffness, derived from the initial slope ($k$) of the force-end extension ($\Delta/L$) curves (**Fig. 4a**), varies more than 3× across configurations, from ~2.8 N/m for 120° to ~9.4 N/m for 60°. Some configurations, such as 60° and 240°, exhibit pronounced stiffening, while the other three states maintain relatively constant stiffness over the same range of extension. This highlights the tunability of stiffness in response to both configuration and applied extension.

This concept naturally extends to stacked assemblies, where combining individually programmable units offers even greater mechanical tunability. We test four combinations of three-unit stacks with independently programmed twist angles (from bottom ($\varphi_1$) to top ($\varphi_3$)), twisted either clockwise (CW) or counterclockwise (CCW): (i) 60°–120°–60° (all CW), (ii) 120°–240°–120° (all CW), (iii) 120°(CCW)–0°–120°(CW), and (iv) 120°(CW)–0°–120°(CW). The mechanical responses vary significantly across these configurations due to differences in sequencing, twist direction (CW vs. CCW), and whether snapping occurs during loading. These differences demonstrate how unit arrangement and handedness offer additional design strategies for programmable mechanical behavior. Experimental results reveal multi-stage stiffness profiles with two or three distinct regimes, marked by $k_1$, $k_2$, and $k_3$, across increasing extension ranges (**Fig. 4b-c**). For instance, $k_1$ of cases (i), (ii), and (iv) is considerably higher than in case (iii), while $k_3$ in (i) and (iii) shows strong stiffening at large extensions. In summary, RodOri introduces programmable tuning of both equilibrium stiffness and nonlinear mechanical behavior, enabling architected structures with reconfigurability and adaptive static properties. These same mechanisms, as shown next, provide a powerful foundation for dynamic functionalities.

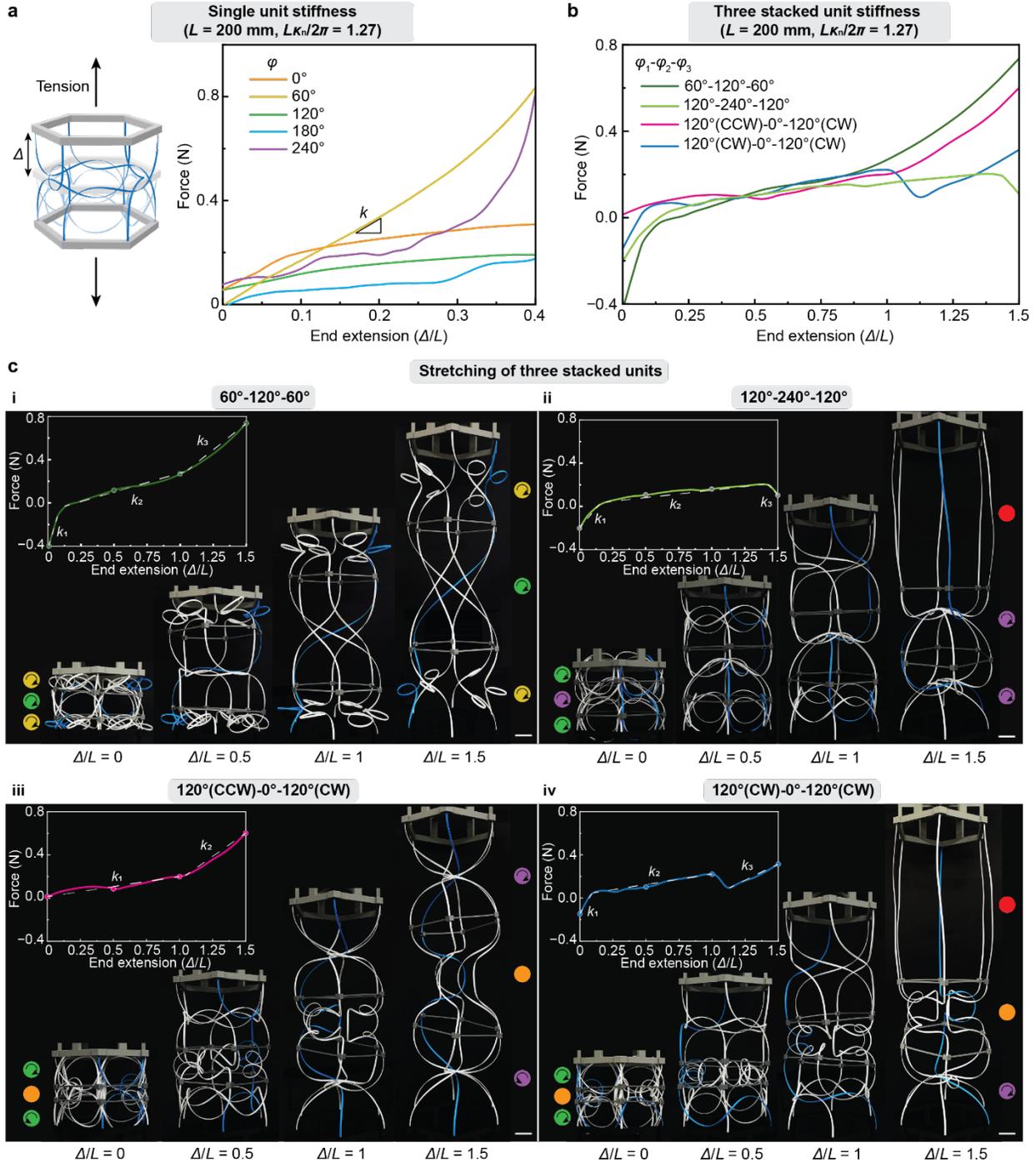

**Fig. 4 | Static stiffness tuning of stacked RodOri. a**, Schematic of a 6-rod RodOri unit in tension. Tensile force vs end extension ($\Delta/L$) for the reconfigured RodOri unit ($L$ = 200 mm, $L\kappa_n/2\pi$ = 1.27, $h/t$ = 4) at 0°, 60°, 120°, 180°, and 240°. **b**, Tensile force vs end extension for an assembly of three stacked 6-rod RodOri units ($L$ = 200 mm, $L\kappa_n/2\pi$ = 1.27, $h/t$ = 4). The initial configuration of the three units is varied, as indicated by the different curves. **c**, Images from the experimental tensile test at end extensions of 0, 0.5, 1, and 1.5 for assemblies with unit configurations of i) 60°–120°–

60° (all CW), ii) 120°–240°–120° (all CW), iii) 120°(CCW)–0°–120°(CW), and iv) 120°(CW)–0°–120°(CW). Scale bars: 2.5 cm.

## 5. Programmable Dynamic Properties of RodOri Metamaterials

The multiple configurations and multistability of RodOri enable active metamaterials with tunable and reprogrammable dynamic mechanical properties. Here, we demonstrate how tessellated assemblies of RodOri enable tunable vibration behavior and wave control through geometric reconfiguration. We investigate a tessellated structure composed of seven 6-rod RodOri units ($L = 200$ mm, $L\kappa_n/2\pi = 1.27$, $h/t = 4$) arranged into a planar metamaterial (**Fig. 5a**). Each unit in the system can be selectively and reversibly transformed from a fully deployed configuration (**Fig. 5a-i**) into any of the five folded configurations demonstrated in **Fig. 3b**. These global configurations can be homogeneous, for example, all units deployed (**Fig. 5a-i**) or all units twisted to 180° (**Fig. 5a-ii**), or heterogeneous, such as a combination of 120° and 180° units (**Fig. 5a-iii**).

To explore dynamic tunability, we examine vibration transmission through the metamaterial in both deployed and folded configurations. A sinusoidal input vibration with amplitude $U_{in}$ is applied at the left base frame of the central unit, and the transmitted amplitude $U_{out}$ is measured at the right base frame, both experimentally and numerically. The vibration transmission, defined as $20 \log(U_{out}/U_{in})$, for the homogeneous metamaterial is plotted as a function of frequency $f$ in **Fig. 5b**. In the homogeneous deployed state, the system behaves as a nearly rigid body with perfect transmission across the frequency range of interest (0 – 15 Hz), i.e., $U_{out}/U_{in} = 1$ (**Fig. 5c-i**). In contrast, folding and selectively programming the folded configuration introduces distributed compliance, dramatically altering the dynamic response. In the homogeneous 180° folded state (**Fig. 5c-ii**), the system exhibits strong frequency-dependent behavior: low-frequency vibrations near a resonance ($f = 2.5$ Hz) are significantly amplified ($U_{out}/U_{in} = 7.8$), while higher frequencies are strongly attenuated ($U_{out}/U_{in} = 0.1$ at $f = 15$ Hz). The resonance frequency can be readily shifted by varying both global configuration of the system and the natural curvature of the constituent rods, owing to their significant influence on the system's effective stiffness. Moreover, in the heterogeneous folded state (**Fig. 5c-iii**), asymmetry in the global configuration leads to symmetry breaking in vibration modes, converting a symmetric mode into an asymmetric one, demonstrating programmed mode conversion. These results confirm that RodOri metamaterials enable a wide range of dynamic functionalities, including amplification, isolation, and mode shaping.

Beyond vibration tuning, we numerically demonstrate dynamic wave switching in a larger metamaterial system of 23 tessellated 6-rod RodOri units ($L = 200$ mm, $L\kappa_n/2\pi = 1.27$, $h/t = 4$), all twisted to 180° (**Fig. 5d**). The bottom base frame is fixed, while the top base frame is free. A localized excitation is applied at the central unit of the top base frame at $f = 50$ Hz. In the deployed state (**Fig. 5d-i**), the wave remains localized near the excitation source, suppressing propagation. In contrast, in the folded state (**Fig. 5d-ii**), the wave propagates broadly across the system, illustrating a clear on-off switching capability. This functionality enables programmable routing of mechanical signal transmission.

As discussed earlier, RodOri with $n$ reconfigurable states tessellated into a metamaterial of $m$ units can produce $n^m$ global configurations and stiffness/elastic energy distributions. Incorporating units made from rods with differing natural curvatures or constituent materials adds another layer of design freedom, further expanding the accessible dynamic response space. This approach enables both coarse and fine control over dynamic mechanical properties, including the operation frequency of the tunable vibration control or wave propagation switch, making it highly suited for adapting to changing environments or functional requirements. In summary, RodOri metamaterials provide a reconfigurable, multistable metamaterial platform for dynamic mechanical control. Through selective and programmable folding, these architected systems achieve tunable vibration filtering, mode conversion, and wave propagation switching, paving the way for multifunctional devices in adaptive vibration isolation, mechanical computing, and wave-based logic systems.

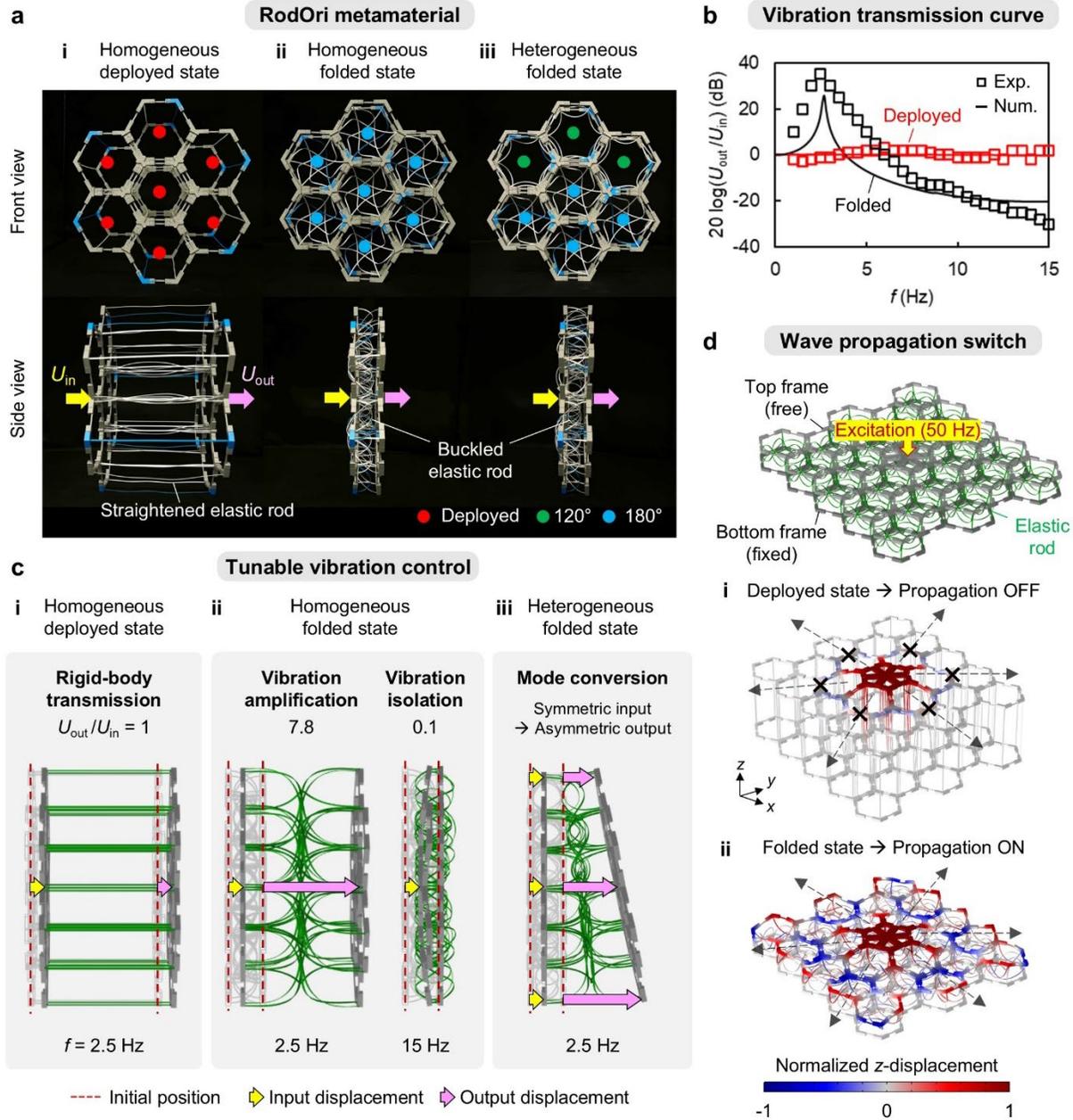

**Fig. 5 | Programmable dynamic properties of RodOri metamaterials. a,** Tessellation of seven 6-rod RodOri units ($L = 200$ mm, $L\kappa_n/2\pi = 1.27$, $h/t = 4$). (i) Homogeneous deployed state; (ii) homogeneous folded state with all units twisted to 180°; (iii) heterogeneous folded state with three units twisted to 120° and four to 180°. **b,** Vibration transmission in the homogeneous deployed state (red) and homogeneous 180° folded state (black), measured as $20 \log(U_{out}/U_{in})$ as a function of frequency $f$, where $U_{in}$ and $U_{out}$ denote the input and output displacement amplitudes at the left and right base frames, respectively. Square markers: experiments; solid lines: numerical simulations. **c,** Tunable vibration control via reconfiguration. (i) Rigid-body transmission in the homogeneous deployed state at 2.5 Hz; (ii) vibration amplification at 2.5 Hz and isolation at 15 Hz in the homogeneous 180° folded state; (iii) mode conversion in the heterogeneous folded state at 2.5 Hz. **d,** On-off mechanical wave propagation switch. The system consists of 23 6-rod RodOri

units ($L = 200$ mm, $L\kappa_n/2\pi = 1.27$, $h/t = 4$) with a fixed bottom base frame and a free top base frame. Colormaps represent the simulated out-of-plane displacements normalized by the excitation amplitude at 50 Hz in the homogeneous (i) deployed and (ii) 180° folded states.

## 6. Conclusions

In summary, we have introduced elastic rod origami (RodOri)—a robust reconfigurable mechanical platform constructed from pre-stressed, naturally curved rods whose stored elastic energy and mechanical instability enable transitions between an unprecedented number of discrete configurations. Upon compression or twisting, RodOri can reversibly transition between different states through elastic deformation-induced snapping. By precisely tailoring rod natural curvature, we design the system's energy landscape, and, in turn, achieve wide tunability of static stiffness within a single RodOri unit at discrete configurations due to the different rod morphologies and interactions among rods. Larger assemblies, formed by stacking or tessellating these units, unlock unprecedented control over global architecture through selective reconfiguration of individual units. This reconfigurability directly translates into programmable control of both equilibrium stiffness and nonlinear mechanical response, influencing the structure's dynamic behavior and enabling adaptive metamaterials capable of tunable vibration filtering, mode conversion, and wave-propagation switching.

The design strategy presented here is readily adaptable to a broad range of materials and geometries, offering a versatile route to mechanical systems with programmable properties. The vast design space—spanning rod curvature and length, cross-sectional geometry, material modulus, and spatial arrangement—remains largely unexplored and promises fine control over both static and dynamic behavior. Embracing heterogeneous designs, exploiting inverse design approaches, and integrating active or stimuli-responsive materials could further extend functionality, enabling on-demand adaptation to rapidly changing environments. We envision RodOri as a broadly applicable framework for engineering multifunctional architected structures, where harnessing mechanical instabilities becomes a deliberate and powerful tool for addressing complex engineering challenges.

## 7. Methods

*RodOri fabrication*

For the RodOri systems, the rods and rigid bases were 3D printed separately from one another and then assembled together. When the dimensionless natural curvature ($L\kappa_n/2\pi$) of the rod exceeded 1, the rod was modeled as a spiral with a small spacing of approximately 1 mm between adjacent coils. All rods were designed with an *h/t* of 4, or height and thickness of 2.2 mm and 0.55 mm, respectively. All parts were printed with polylactic acid (PLA Basic, Bambu Lab) using an X1 Carbon (Bambu Lab) 3D printer.

*Structural simulation*

The commercial software ABAQUS 2024 (Dassault Systèmes, France) was used to simulate the deformation of the RodOri systems. The rods with non-zero natural curvature were modeled as straight, thermally stressed bilayers. The RodOri was subsequently loaded to trigger buckling, and penetration between rods was allowed. C3D8R elements were used for all models.

*Mechanical testing*

All mechanical testing of the rods and RodOri was performed via displacement-controlled tests using a universal testing machine (3344, Instron, Inc., USA). For the compression testing of the rods (**Fig. 2a**), the rods with natural curvature were straightened and then clamped into the testing machine at either end and compressed then unloaded at a rate of $0.01L$ s$^{-1}$, where $L$ is the length of the rod. For the tensile testing of the RodOri units (**Fig. 4**), the RodOri in its stable configuration was clamped into the testing machine at either end and stretched at a rate of $0.01L$ s$^{-1}$.

*Vibration transmission experiment*

To experimentally characterize the dynamic response of the tessellated RodOri, the system was hung on a test bench using rubber bands, with the central unit of the input frame connected to a modal shaker (Model JZK-2, Sinocera Piezotronics, China). Accelerometers (Model 352A21, PCB Piezotronics, USA) were installed on the central units of the input and output frames to measure vibrational displacements.

*Vibration transmission simulation*

To numerically evaluate the dynamic response of the tessellated RodOri, FEA simulations were performed using the Solid Mechanics module in COMSOL Multiphysics 6.1. The 3D-printed elastic rods were modeled as a linear elastic solid, with a Young's modulus of 2.6 GPa, mass density of 1.24 kg m$^{-3}$, and Poisson's ratio of 0.33. Structural damping was incorporated via a loss factor of 0.05. Harmonic excitation was applied through a prescribed displacement at the central unit of the input frame, while all other boundaries were set to traction-free.

*Wave propagation simulation*

To investigate the wave propagation behavior of the tessellated RodOri, FEA simulations were performed using the Solid Mechanics module in COMSOL Multiphysics 6.1. The bottom frame was fully constrained using fixed boundary conditions, while the top frame was traction-free to allow out-of-plane motion. Harmonic excitation was introduced by a prescribed displacement at the central unit of the top frame.